# Polarization properties of laser-diode-pumped micro-grained Nd:YAG ceramic lasers

Kenju Otsuka and Takayuki Ohtomo

*Department of Human and Information Science, Tokai University, 1117 Kitakaname, Hiratsuka, Kanagawa 259-1292, Japan*

**Abstract:** Detailed polarization properties have been examined in laser-diode-pumped (LD-pumped) micro-grained ceramic Nd:YAG lasers in different microchip cavity configurations. Stable linearly-polarized single-frequency oscillations, whose polarization direction coincide with that of an LD pump light, were observed in an external cavity scheme. While, in the case of a thin-slice laser scheme with coated reflective ends, elliptically-polarized single-frequency operations took place in the low pump-power regime and dynamic instabilities appeared, featuring self-induced antiphase modulations among counter-rotating circularly-polarized components having slightly different lasing frequencies, with increasing the pump power.

The polarization properties of radiation from single-crystalline Nd:YAG lasers have been studied from various aspects, including such effects as the absorption and amplification anisotropy due to the local symmetry of Nd ions in the YAG crystal matrix [1, 2] an anisotropy of the laser resonator [3] and a depolarization due to the thermal birefringence, particularly in high-power operations [4]. The polarization state, in general, depends on many parameters and the study of polarization properties of Nd:YAG lasers is of interest for analysis of the properties of the so-called vector lasers and for practical applications similar to vertical cavity surface emitting semiconductor lasers (VCSELs) [5]. Different polarization properties have been reported experimentally in LD-pumped microchip Nd:YAG single crystalline lasers. These include linear polarization oscillations along the polarization direction of the LD pump light in the case of an isotropic laser cavity [6] and orthogonal linear polarization (i.e., dual-polarization) oscillations in the case of quasi-anisotropic laser cavity [3, 7]. As for Nd:YAG ceramic lasers that consist of small single-crystalline grains with distributed crystal axes, on the other hand, polarization properties have been demonstrated to depend on the average grain size with respect to the lasing beam diameters [7]. We have reported $TEM_{00}$-mode oscillations in micro-grained Nd:YAG and Yb:YAG ceramic lasers whose average grain size is smaller than about 5 µm, resulting from spatial averaging of thermal birefringence and the pronounced stress relaxation at increased grain boundaries [7], without segregations into local modes which take place in larger grain ceramic lasers of several tens of microns [8].

In this paper, we re-examined polarization properties of LD-pumped micro-grained Nd:YAG ceramic lasers, paying special attentions to dependences of polarization of radiation on the pump polarization and resonator isotropy. To study the effect of resonator isotropy, we employ a semi-confocal external cavity which supports well-defined $TEM_{00}$ transverse eigenmode and a plane-parallel thin-slice laser with coated reflective ends whose resonator configuration is determined by the thermal lens effect and the LD pump-beam (gain) profile, in which a weak resonator anisotropy is expected.

The experiment was carried out by using 1-mm-thick Nd:YAG ceramic samples with Nd concentration of 1 at.% purchased from the same company (Baikowski Japan) in different cavity configurations as shown in Fig. 1(a) and (b). In both cases, a nearly collimated elliptical LD beam with a wavelength of 808 nm was transformed into a circular one by an anamorphic prism pair and focused onto the ceramic samples by using a microscope objective lens with a numerical aperture of 0.25. The LD pump light was linearly polarized along the 1-µm thick, 100-µm wide active layer (i.e., x-axis depicted in Fig. 1). The pump-beam spot size averaged over the 1-mm-thick sample length was measured to be 80 μm, where the absorption coefficient for the LD light was 3.55 cm$^{-1}$. In the case of Fig. 1(a), the sample was placed within a semi-confocal external cavity, in which the sample was attached to a plane mirror $M_1$ (99.8% reflective at 1064 nm and >95% transmissive at 808 nm) and a concave mirror $M_2$ (99% reflective at 1064 nm, radius of curvature: 10 mm) was placed 5 mm away from the plane mirror. In the case of Fig. 1(b), the end surfaces of the Nd:YAG ceramic sample were coated with dielectric mirrors which have the same coating properties as $M_1$ and $M_2$ in Fig. 1(a). We have tested several samples and the following results have been commonly obtained.

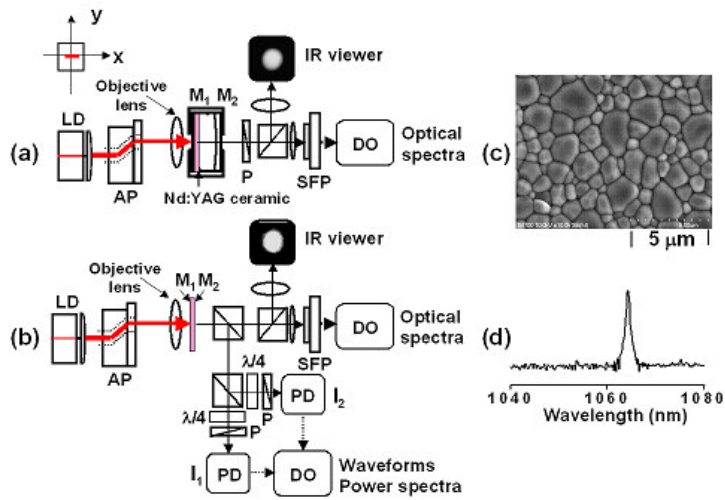

Fig. 1. Experimental setup of LD-pumped Nd:YAG microchip ceramic lasers. (a) Semi–confocal cavity configuration. (b) Mirror-coated thin-slice cavity configuration. LD: laser diode, AP: anamorphic prism, P: polarizer, SFP: scanning Fabry-Perot interferometer, λ/4: quarter-wave plate, PD: photo-diode, DO: digital scilloscope. We used polarization-independent beam-splitters. (c) Typical SEM image of the thermally-etched surface of the ceramic samples. (d) Global oscillation spectrum.

A typical surface SEM (scanning electron microscope) image of the thermally-etched surface of the samples we used for both cavity configurations is depicted in Fig. 1(c), where the average grain size was determined to be 1.1 μm by averaging the length along the longest direction of each grain [7]. Single-transition oscillations on the 1064-nm transition line were confirmed for the entire pump-power region in both lasers by measurements using a multi-wavelength meter as shown in Fig. 1(d). Input-output characteristics of both lasers are shown in Figs. 2 and 3, respectively. The threshold pump power was lower for the external cavity laser presumably because of a better match between the pump and lasing beams. Far-field patterns and the corresponding detailed oscillation spectra were measured with a PbS infrared viewer and a scanning Fabry-Perot interferometer of a 10-MHz frequency resolution. Segregations into local modes [8] were not observed and the single longitudinal- and transverse-mode oscillations were maintained in the entire pump-power region in both lasers. Typical far-field patterns are shown in the insets of Figs. 2 and 3.

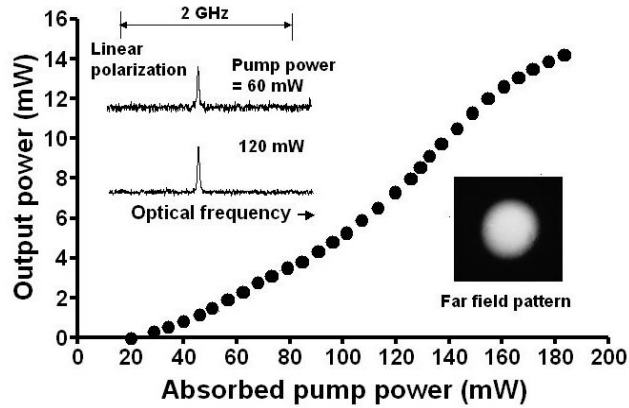

Fig. 2. Input-output characteristic, oscillation spectra and near-field pattern for the external cavity Nd:YAG ceramic laser.

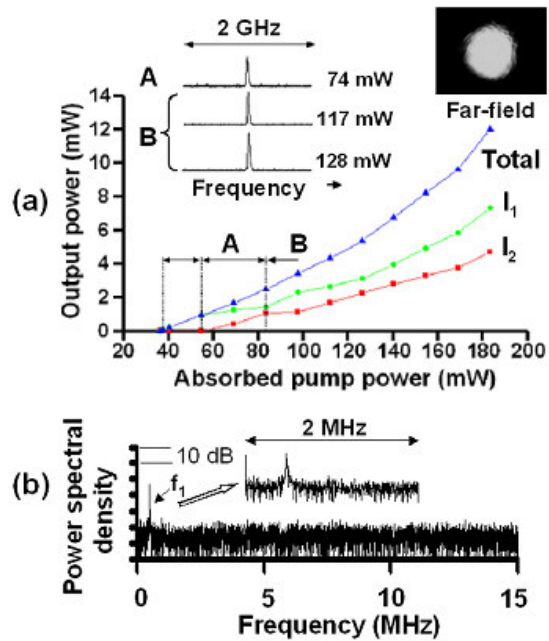

Fig. 3. (a) Polarization-dependent input-output characteristics, optical spectra and near-field pattern for the mirror-coated 1-mm-thick Nd:YAG ceramic laser. (b) Output power spectrum in region **A**.

As for polarization of radiations, two lasers exhibited different properties. In the external cavity laser, linearly-polarized single-frequency oscillations whose direction coincided with that of pump polarization were obtained similar to the Nd:YAG single-crystalline laser in the isotropic resonator [6]. Pump-dependent oscillation spectra are depicted in the inset of Fig. 2. In the case of the thin-slice laser, on the other hand, the radiation polarization state was found to depend on the pump power although the monomode operation was maintained as shown in the inset of Fig. 3(a). In the low pump-power region indicated by the arrow, linearly polarized single-frequency emissions along the pump polarization direction took place similar to the external cavity laser. With increasing the pump power, however, the linear polarization state was found to be replaced by elliptically-polarized single-frequency oscillations in the limited pump-power region **A** indicated in Fig. 3(a).

We have separated counter-rotating circularly-polarized emissions into orthogonal linear polarization field components which are at 45° with x- and y-axes, $E_1$, $E_2$, by using circular-polarization filters consisting of a quarter-wave plate and a linear polarizer as depicted in Fig. 1(b) and individual output powers $I_{1,2} = |E_{1,2}|^2$ are indicated in Fig. 3(a). As the pump power was increased further into region **B**, the "frequency-locking" of two counter-rotating circularly-polarized components failed and two components belonging to the same longitudinal mode were found to oscillate at slightly different frequencies which cannot be well resolved by the scanning Fabry-Perot interferometer we used. Pump-dependent oscillation spectra are shown in the inset of Fig. 3(a). A similar bistability of the left and right quasi-circularly polarized laser fields oscillating at slightly different optical frequencies, with a phase difference of π, has been predicted for a quasi-isotropic longitudinally monomode Nd:YAG single-crystalline laser [9].

The observed polarization property of the Nd:YAG micro-grained ceramic laser in the isotropic laser cavity is similar to that of the Nd:YAG single-crystalline laser [6]. This result suggests that the anisotropy of amplification is caused by the angular distribution of population inversions resulting from the saturation of the population inversions by the polarized radiation rather than the local electric field of compensating charges of Nd ions in the local $D_2$ (orthorhombic) symmetry [1,2] because crystal axes are randomly distributed in micro-grained ceramic samples. The angular inhomogeneity depends on the orientation of the polarization plane of laser radiation from that of pump radiation and the polarization state is considered to be almost completely determined by polarization of pump radiation for the isotropic cavity [6,12].

As for the thin-slice mirror-coated micro-grained Nd:YAG ceramic laser, which is considered to possess a pump-beam-dependent weak cavity anisotropy, quasi-elliptically polarized emissions, consisting of counter-rotating circularly-polarized modes with slightly different optical frequencies, become dominant with the increased pump-power region **B** as shown in Fig. 3(a). To verify the speculated pump-power dependent frequency-locking and its failure among two counter-rotating circularly-polarized emission components experimentally, dynamic properties of two components were examined. In region **A** of Fig. 3(a), in-phase relaxation oscillations driven by a white noise appear in two components and only single noise peak appears at the relaxation oscillation frequency, $f_1 = (1/2\pi)[(w - 1)/\tau\tau_p]^{1/2}$ ($w = P/P_{th}$, $P_{th}$: threshold pump power, $\tau$: fluorescence lifetime, $\tau_p$: photon lifetime) as shown in Fig. 3(b). This implies that two components are comprised in the single-frequency elliptically polarized lasing mode. In region **B**, on the other hand, two components are found to exhibit non-identical outputs modulated at the frequency $f_B > f_1$. The modulation frequency was found to decrease with increasing the pump power. Typical modal intensity waveforms and the corresponding power spectra in region **B** in Fig. 3(a) are shown in Fig. 4, together with those for the total output. It is obvious that two components behave as different lasing modes in this region featuring clear antiphase modulation dynamics at $f_B$ as indicated by the arrows in magnified intensity waveforms. The modulation frequency component in the power spectrum of the total output is strongly suppressed by about -20 dB. Similar quasi-elliptically polarized emissions and the states of modulated polarization have been theoretically predicted to occur in the model of VCSELs with a small loss anisotropy [5].

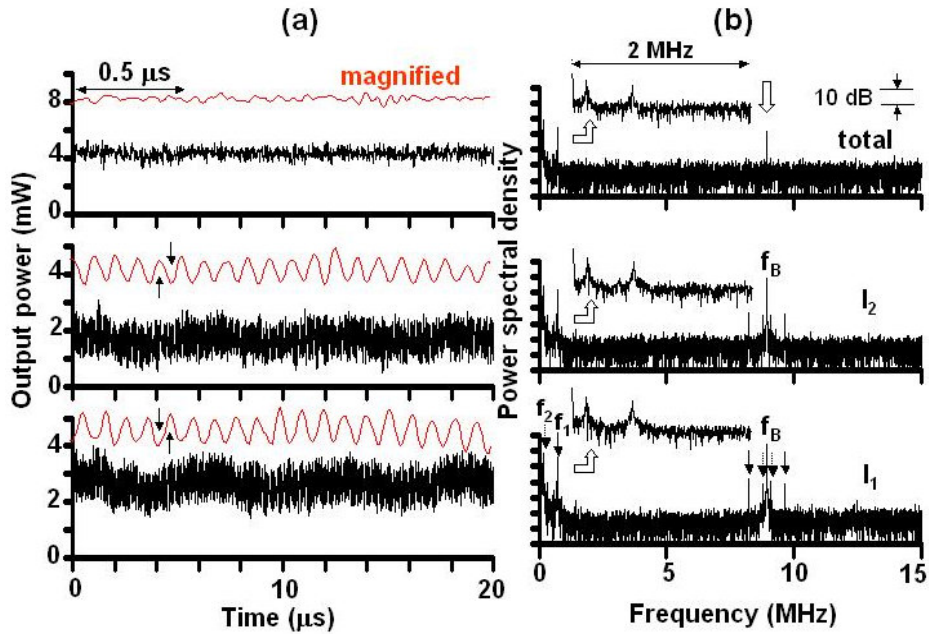

Fig. 4. (a) Modal and total output intensity waveforms and (b) power spectra observed in the mirror-coated 1-mm-thick Nd:YAG ceramic laser operating in region **B** of Fig. 3. Pump power = 117 mW.

Intensity components polarized along x- and y-axes, $I_x$ and $I_y$, were measured by removing quarter-wave plates and setting polarizers along x and y directions in Fig.

1(b). A typical example is shown in Fig. 5. Both components exhibited modulated outputs, however, the antiphase dynamics was violated, where the modulation component in the total output was always strongly suppressed in region **B** similar to Fig. 4(a). Therefore, these components ($I_x$, $I_y$) are not lasing eigen-polarization modes and the simple orthogonal linearly polarized emission is ruled out.

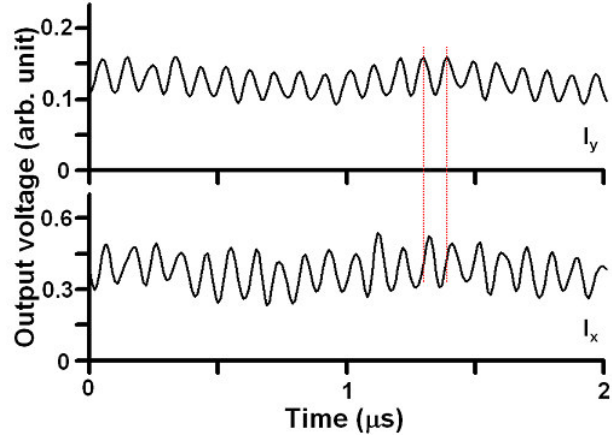

Fig. 5. Intensity waveforms polarized along x and y directions in region **B** of Fig. 3. Pump power = 120 mW.

On the other hand, self-induced modulations at the beat frequency between orthogonal linear polarization modes have been reported in a Nd:YAG single-crystalline laser, which includes intracavity optical elements ensuring a longitudinally monomode oscillation and providing a loss as well as a phase anisotropy with a 38.5-cm long semi-spherical cavity [3]. However, output intensities of orthogonal linear polarization modes, $I_x$, $I_y$, were reported to show in-phase modulations, while the antiphase relaxation oscillation dynamics at very low frequencies have been reported to hold [3]. The inherent antiphase relaxation oscillation dynamics in multimode solid-state lasers operating in a single polarization, i.e. scalar lasers [10, 11], which arises from the cross-saturation of population inversions among different longitudinal modes through the spatial hole-burning effect, is violated in the present micro-grained Nd:YAG ceramic laser. In short, if we look at power spectra in the low frequency region, another relaxation oscillation noise component appears at $f_2$ ($< f_1$), suggesting two mode oscillations, i.e., the failure of frequency-locking, but this lower-frequency component does not vanish for the total output unlike the well-known antiphase dynamics in the model of scalar multimode lasers [9, 10]. Note that relaxation-oscillation-induced sidebands appear at $f_B \pm f_1$ and $f_B \pm f_2$ as shown in the modal power spectra. The angular inhomogeneity of population inversions, i.e., vector nature, could also be attributed to the observed behavior because the spatial hole-burning along the longitudinal direction is absent in for circularly-polarized lasing eigenmodes [12]. The most conventional mirror-coated laser scheme is also considered to be suitable for micro-grained Nd:YAG ceramics to realize coherent compact light sources of high modal purity because the total output is stabilized owing to the antiphase modulation dynamics.

In summary, polarization properties of radiations from LD-pumped micro-grained Nd:YAG ceramic lasers have been studied experimentally. Lasing eigen-polarization states in microchip lasers are shown to depend on the isotropy of laser cavities. In the isotropic external cavity configuration, single-frequency oscillations were observed in a linearly polarized state whose polarization direction coincides with that of pump light. In the mirror-coated thin-slice laser configuration with a pump-beam dependent weak cavity anisotropy, elliptically-polarized single and closely-separated two-frequency quasi-elliptically polarized operations, which feature antiphase modulation dynamics, were observed. Stable single longitudinal and transverse-mode operations are found to be attained with LD-pumped micro-grained Nd:YAG ceramic lasers independently of microchip cavity configurations. Detailed theoretical studies on observed curious polarization properties in micro-grained Nd:YAG ceramic lasers are strongly anticipated from the viewpoint of nonlinear dynamics of vector lasers.